# EOM Minimum Point Bias Voltage Estimation for Application in Quantum Computing

Frank Obernosterer **, Raimund Meyer **, Robert Koch *,
Gerd Kilian *, Ewald Hedrich *, Christian Kelm *

* *Fraunhofer Institute for Integrated Circuits IIS, Erlangen, Germany*
*POC: robert.koch@iis.fraunhofer.de*

** *Com-Research GmbH, Solutions for Communication Systems, Fürth, Germany*

## 1 Introduction

In quantum computing systems the quantum states of qubits can be modified among others by applying light pulses. In order to achieve low computing error rates these pulses have to be precisely shaped in magnitude and phase. A widely used approach is to modulate or switch the light of a stabilized continuous wave (CW) laser by a voltage controlled modulator. In practical applications, both acousto-optic (AOM) and electro-optic modulators (EOM) are used for this purpose. There are specific advantages and disadvantages to each of these technologies. The advantages of EOMs, in particular Mach-Zehnder modulators (MZMs), include e.g. higher bandwidth [1],[2], compactness, good integrability [3], and better noise performance [4]. On the other hand, EOMs are challenging regarding their control voltage regulation as they experience a bias drift, i.e. voltage shifts in the modulator's operating point. This is caused by temperature fluctuations and long-term changes in material properties, which alter the refractive index [1],[2]. In addition, applying a DC control voltage leads to charge accumulation in the electro-optic material, causing changes in the internal electric fields. These shifts can alter the refractive index of the material, which also leads to a drift in the modulator’s operating point and a corresponding shift in the transfer function [5]. Such a shift significantly impairs the quality of the modulation, which is why EOMs usually require a bias control loop to track the DC operating point.

This work addresses the estimation of the instantaneous bias voltage using a small pilot tone, where the optical output power of the EOM is tracked by a photodetector feedback signal.

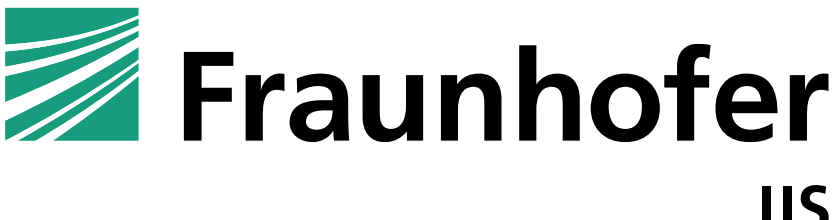

## 2 Background

The most common bias control methods for Mach-Zehnder modulators mainly include feedback control, based on the ratio between optical input power and optical output power (so-called pilot-free methods) [8] or feedback control techniques based on harmonic pilot signals [9,10] injected into the bias port of the modulator. In the pilot signal control methods, either the minimum ratio between the first and second order harmonics (at minimum or maximum point) or between the second and first order harmonics (at positive or negative quadrature points) is being searched for the desired operating point in the EOM transfer function [9]. Alternatively, a technique can be used, that utilizes the optical output power and first-order harmonic of the pilot to lock the EOM to the different operating points of the transfer function [10]. All these mentioned pilot signal control methods use a second- or fourth-order Taylor series expansion to approximate the wanted operating points of the transfer function.

Here we focus on biasing the EOM to operate at its minimum output point, which is relevant for the time intervals between optical pulses or pulse trains in quantum computing applications. We extend the existing approaches by taking the phase of the pilot signal into account or, equivalently, the timing between the applied pilot signal and the corresponding EOM output feedback signal. This allows to provide immediate, memoryless information about the voltage offset between the current operation point and the minimum output point. The algorithm proposed here not only estimates this voltage offset in terms of magnitude but also includes information on the sign. This achieves significantly better convergence in tracking, especially if the feedback signal is very noisy compared to the useful signal contributions. The scope of this paper is not on the actual control loop, but on the estimation of the deviation signal serving as input to the control loop.

The paper is organized as follows: In section 3, the estimation of the voltage offset between the current operation point and the minimum output point is derived theoretically, based on the widely used transfer function model for the EOM [6]. We include the phase of the applied sinusoidal pilot signal as a parameter. Section 4 considers aspects of a real-world system, such as the key parameters of a photodetector, the sampling of its output signal and the propagation delay between control signal and feedback signal. Furthermore, the variance of the offset estimation error is determined by calculation depending on the relevant system parameters. Section 5 compares the mathematically derived estimation error variance, which is based on certain simplifying model assumptions, with corresponding simulation results without these assumptions. Furthermore, results are shown how the pilot signal itself raises the minimum achievable output power of the EOM. Finally, design criteria for pilot signal parameters such as amplitude, frequency, duration, etc. are listed and discussed at a qualitative level.

## 3 Theoretical Analysis

### 3.1 Approximation of the transfer function in the vicinity of its minimum

The basic EOM transfer function from the applied control voltage $V_c(t)$ to the output optical power $P_{out}$, including the optical imbalance factor $f_{ib} \leq 0.5$ between the two branches of the MZM, is given by [7]:

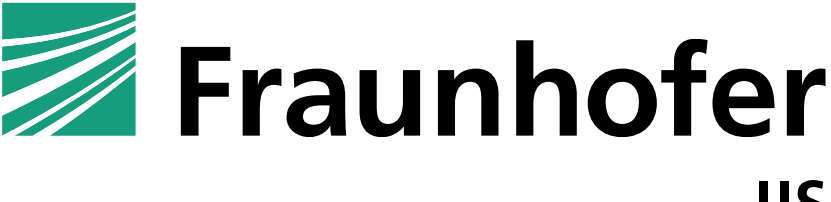

$$P_{out}(t) = P_{in} \cdot \left( \frac{1}{2} + f_{ib} \cdot cos \left( \frac{\pi}{V_\pi} \cdot \left( V_0 - V_c(t) \right) \right) \right). \tag{1}$$

The value $\varphi_0 = V_0 \frac{\pi}{V_\pi}$ characterizes the initial phase shift that would exist without the presence of an electric field. $V_\pi$ is the half-wave voltage in order to drive the optical power from the maximum value to the minimum value (or vice versa) and $V_c(t)$ is the control voltage that generates an electric field between the two electrodes. $P_{in}$ denotes the input optical power, which is assumed to be constant.

Starting from the current operating point of the EOM with the bias voltage $V_{DC} = \hat{V}_{min}$, the voltage $V_{min}$ is searched that leads to the actual minimum possible optical output power of the EOM according to Figure 1b. The (previously) estimated value $\hat{V}_{min}$ deviates from the exact value $V_{min}$ by the differential voltage $\Delta V$ to be estimated, according to the relationship

$$\Delta V = V_{min} - \hat{V}_{min}. \tag{2}$$

Accordingly, the knowledge of the differential voltage $\Delta V$ allows the calculation of the required voltage $V_{min}$. In the following, the procedure for estimating $\Delta V$ is derived, which is directly accompanied by an estimate of $V_{min}$.

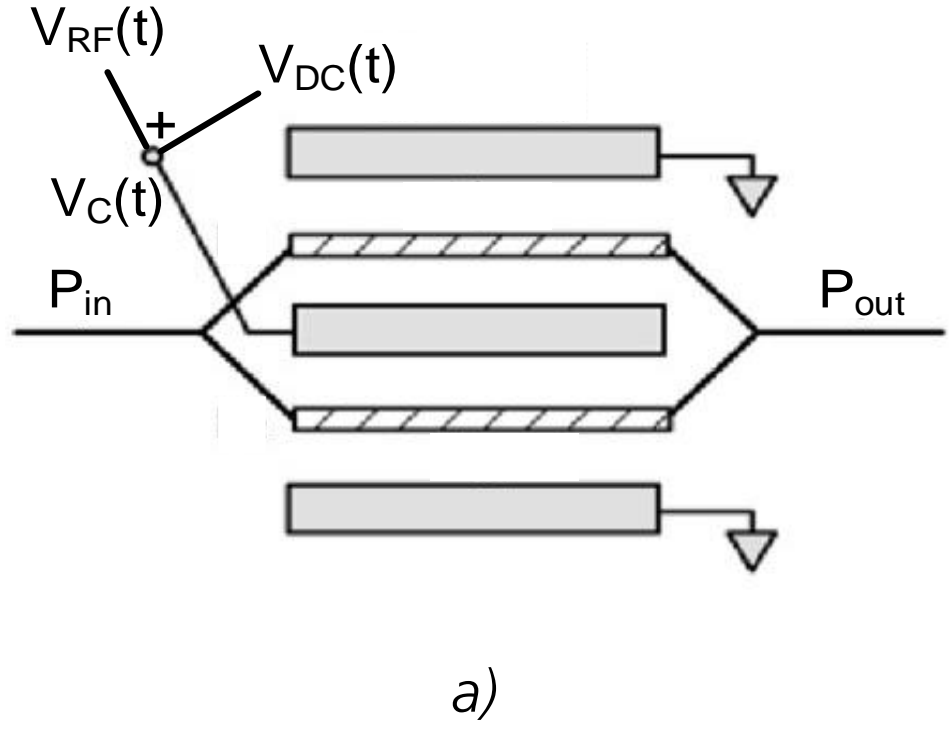

*a)*

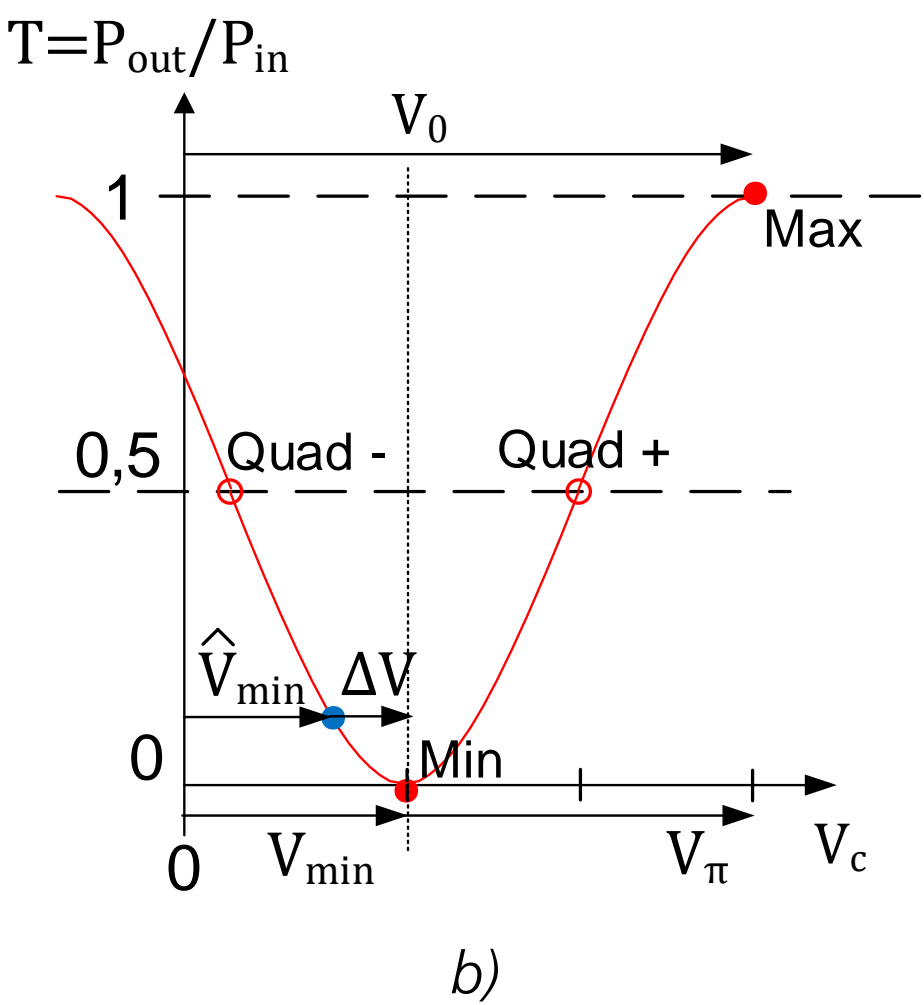

*b)*

*Figure 1: a) Schematic geometric illustration and b) transfer function curve of a MZM*

The transfer function model according to Eq. (1) is periodic and therefore has an infinite number of minima according. To simplify the following illustration, we refer to the minimum at $V_{min} = V_0 - V_\pi$ without restricting the general validity, as shown in Figure 1b.

If the EOM is biased at the DC operating point $V_{DC} = \hat{V}_{min}$ as assumed in Figure 1b and a sinusoidal pilot signal

$$V_{RF}(t) = V_{pilot}(t) = -V_d \cdot sin(\omega_d t + \varphi_d) \tag{3}$$

with the angular frequency $\omega_d = 2\pi f_d$ and the initial phase $\varphi_d$ is superimposed on this bias, the total control signal is given by

$$V_c(t) = \hat{V}_{min} - V_d \cdot sin(\omega_d t + \varphi_d) = V_0 - V_\pi - \Delta V - V_d \cdot sin(\omega_d t + \varphi_d). \tag{4}$$

The initial phase $\varphi_d$ describes the phase position of the sinusoidal pilot signal relative to a definable time $t = 0$. It is made up of a freely selectable phase $\varphi_0$ and a runtime-related phase difference $\Delta\varphi$, which will be discussed in more detail later:

$$\varphi_d = \varphi_0 + \Delta\varphi. \tag{5}$$

After inserting $V_c(t)$ from Eq. (4) into Eq. (1), we obtain for the optical output power $P_{out}$:

$$P_{out} = P_{in} \cdot \left( \frac{1}{2} + f_{ib} \cdot cos\left( \pi + \frac{\pi}{V_\pi} \cdot \left( \Delta V + V_d \cdot sin(\omega_d t + \varphi_d) \right) \right) \right). \tag{6}$$

In the vicinity of its minimum at $\pi$, i.e. at its development point $x \approx \pi$ and correspondingly for $|\Delta V|/V_\pi \ll 1$ and $V_d/V_\pi \ll 1$, the trigonometric function $\cos(x)$ can be approximated with high accuracy using the quadratic Taylor series according to:

$$\cos(x) \approx -1 + \frac{1}{2} \cdot (x - \pi)^2. \tag{7}$$

Replacing the cosine in Eq. (6) by the quadratic polynomial, $P_{out}$ can be well approximated by $P_{out,T_2}$:

$$P_{out} \approx P_{out,T_2} = P_{in} \cdot \left( \frac{1}{2} + f_{ib} \cdot \left( -1 + \frac{1}{2} \left( \frac{\pi}{V_\pi} \right)^2 \cdot \left( \Delta V + V_d \cdot sin(\omega_d t + \varphi_d) \right)^2 \right) \right). \tag{8}$$

By applying the binomial formula to the quadratic term $\left( \Delta V + V_d \cdot sin(\omega_d t + \varphi_d) \right)^2$, the result

$$\begin{aligned} &\left( \Delta V + V_d \cdot sin(\omega_d t + \varphi_d) \right)^2 = \\ &= \frac{1}{2}(V_d)^2 + (\Delta V)^2 + 2\Delta V \cdot V_d \cdot sin(\omega_d t + \varphi_d) + \frac{1}{2}(V_d)^2 \cdot sin\left( 2\omega_d t + 2\varphi_d - \frac{\pi}{2} \right) \end{aligned} \tag{9}$$

is obtained after a few rearrangements.

## 3.2 Frequency components of the optical output power

If we consider the spectral components of the output signal $P_{out,T_2}$ from Eqs. (8) and (9), we obtain contributions at three frequencies: $P_{out,T_2}(\omega = 0)$, $P_{out,T_2}(\omega = \omega_d)$ und $P_{out,T_2}(\omega = 2\omega_d)$. The real-valued scaling factors $S_{*,T_2}$, which represent the magnitudes of the three spectral components, and the associated phases $\varphi_{*,T_2}$ are:

- DC component at $\omega = 0$: $S_{DC,T_2} = P_{in} \cdot \left(\frac{1}{2} - f_{ib} + \frac{f_{ib}}{2}\left(\frac{\pi}{V_\pi}\right)^2 \left(\frac{1}{2}(V_d)^2 + (\Delta V)^2\right)\right)$.
- 1st-order $(1f_d)$ harmonic at $\omega = \omega_d$: $S_{1f_d,T_2} = P_{in} \cdot \frac{f_{ib}}{2}\left(\frac{\pi}{V_\pi}\right)^2 \cdot 2 \cdot \Delta V \cdot V_d$
  with phase offset $\varphi_{1f_d,T_2} = \varphi_d$.
- 2nd-order $(2f_d)$ harmonic at $\omega = 2\omega_d$: $S_{2f_d,T_2} = P_{in} \cdot \frac{f_{ib}}{2}\left(\frac{\pi}{V_\pi}\right)^2 \cdot \frac{1}{2}(V_d)^2$
  with phase offset $\varphi_{2f_d,T_2} = 2\varphi_d - \frac{\pi}{2}$.

Both the DC and the 1st-order harmonic component $(S_{DC,T_2}, S_{1f_d,T_2})$ depend on the variable $\Delta V$ being searched for and estimated, while the 2nd-order harmonic component $S_{2f_d,T_2}$ is independent of $\Delta V$. There are multiple ways to estimate the required $\Delta V$. In order to become as independent as possible of physically determined variables such as $P_{in}$, $f_{ib}$, the 1st- and 2nd-order harmonic components of $P_{out,T_2}$ are set in relation to each other to determine $\Delta V$, taking amplitude and phase information as well as the phase angle $\varphi_d$ of the pilot signal into account.

## 3.3 Computation of ΔV based on 1st- and 2nd-order harmonics

The Fourier transform $\boldsymbol{\mathcal{F}}\{P_{out,T_2}(t)\}$ of the 1st- and 2nd-order components of $P_{out,T_2}$ of Eq. (8) is considered. We apply the following definition of the Fourier transform of a function $x(t)$:

$$\boldsymbol{\mathcal{F}}\{x\} = \int_{-\infty}^{+\infty} x(t) \cdot e^{-i\omega t}\, dt.$$

As the DC component is irrelevant for the following considerations, it is for reasons of clarity not taken into account further. As the output power $P_{out} \approx P_{out,T_2}$ is a real-valued signal, negative frequencies are also not considered here. This is expressed by the specific Fourier operator $\boldsymbol{\mathcal{F}}^{+}_{1f2f}$. For the Fourier transform of the 1st- and 2nd-order components of $P_{out,T_2}$ at positive values of $\omega$ we obtain:

$$\boldsymbol{\mathcal{F}}\{P_{out,T_2}(t)\}(\omega > 0) =$$

$$\begin{aligned} &= \boldsymbol{\mathcal{F}}^{+}_{1f2f}\left\{S_{1f_d,T_2} \cdot sin(\omega_d t + \varphi_d) + S_{2f_d,T_2} \cdot sin\left(2\omega_d t + 2\varphi_d - \frac{\pi}{2}\right)\right\}(\omega) = \\ &= -S_{1f_d,T_2} \cdot i \cdot \pi \cdot e^{i\varphi_d} \cdot \delta(\omega - \omega_d) - S_{2f_d,T_2} \cdot \pi \cdot e^{i2\varphi_d} \cdot \delta(\omega - 2\omega_d). \end{aligned} \tag{10}$$

If the values of the above Fourier transform at $\omega = \omega_d$ and $\omega = 2\omega_d$ are set in relation to each other, the result after appropriate reductions is:

$$\frac{\mathcal{F}^{+}_{1f2f}(\omega_d)}{\mathcal{F}^{+}_{1f2f}(2\omega_d)} = \frac{-S_{1f_d,T_2} \cdot i \cdot \pi \cdot e^{i\varphi_d}}{-S_{2f_d,T_2} \cdot \pi \cdot e^{i2\varphi_d}} = \frac{S_{1f_d,T_2} \cdot e^{i\left(\varphi_d + \frac{\pi}{2}\right)}}{S_{2f_d,T_2} \cdot e^{i2\varphi_d}} = 4\frac{\Delta V}{V_d} \cdot e^{i\left(-\varphi_d + \frac{\pi}{2}\right)}. \quad (11)$$

After rearranging, we obtain for $\Delta V$:

$$\Delta V = \frac{\mathcal{F}^{+}_{1f2f}(\omega_d)}{\mathcal{F}^{+}_{1f2f}(2\omega_d)} \cdot \frac{V_d}{4} \cdot e^{i\left(\varphi_d - \frac{\pi}{2}\right)}. \quad (12)$$

If the phase offset of the pilot signal $\varphi_d$ and its amplitude $V_d$ are known, the value of $\Delta V$ can be determined both in magnitude and sign simply from the quotient of the Fourier transform of the output power $P_{out} \approx P_{out,T_2}$ at $\omega = \omega_d$ and $\omega = 2\omega_d$. The computation of $\Delta V$ according to Eq. (12) requires neither physical EOM parameters like e.g. $V_\pi, f_{ib}$ nor absolute EOM output power values.

## 4 Aspects of estimating ΔV in a practical system

Eq. (12) describes a comparatively simple way to compute $\Delta V$ based on a small pilot signal by computing the ratio of two spectral components of the EOM output power $P_{out}$ at $\omega = \omega_d$ and $\omega = 2\omega_d$. In a practical implementation, the actual EOM output power $\hat{P}_{out}$ is typically measured by a photodetector (e. g. a photodiode), converting the optical output power into a voltage, which is amplified, filtered and sampled by an ADC.

In the following we further develop the above theoretical approach of computing $\Delta V$, so far based on the knowledge of the time-continuous output $P_{out}(t), V_d, \varphi_d$, to estimate $\Delta V$ in a practical system. We specifically address the estimation error of $\Delta V$ in the presence of photodetector noise, thus allowing to design magnitude and duration of the pilot signal.

### 4.1 Determining phase angle $\boldsymbol{\varphi_d}$

For the computation of $\Delta V$ knowledge of the phase angle $\varphi_d = \varphi_0 + \Delta\varphi$, see Eq. (5), is required. This describes the effective phase shift of the sinusoidal pilot voltage $V_{pilot}(t)$ relative to the time $t = 0$. This temporal zero point can in principle be chosen arbitrarily, but then applies jointly for the pilot signal and the output power $P_{out}(t)$.

Figure 2 shows a pilot signal according to Eq. (3) with $\varphi_0 = 0$, which was chosen to be identical to zero for $t < 0$, i.e. $t = 0$ refers to the start of the pilot signal. In addition, a hypothetical curve of the associated output power is shown.

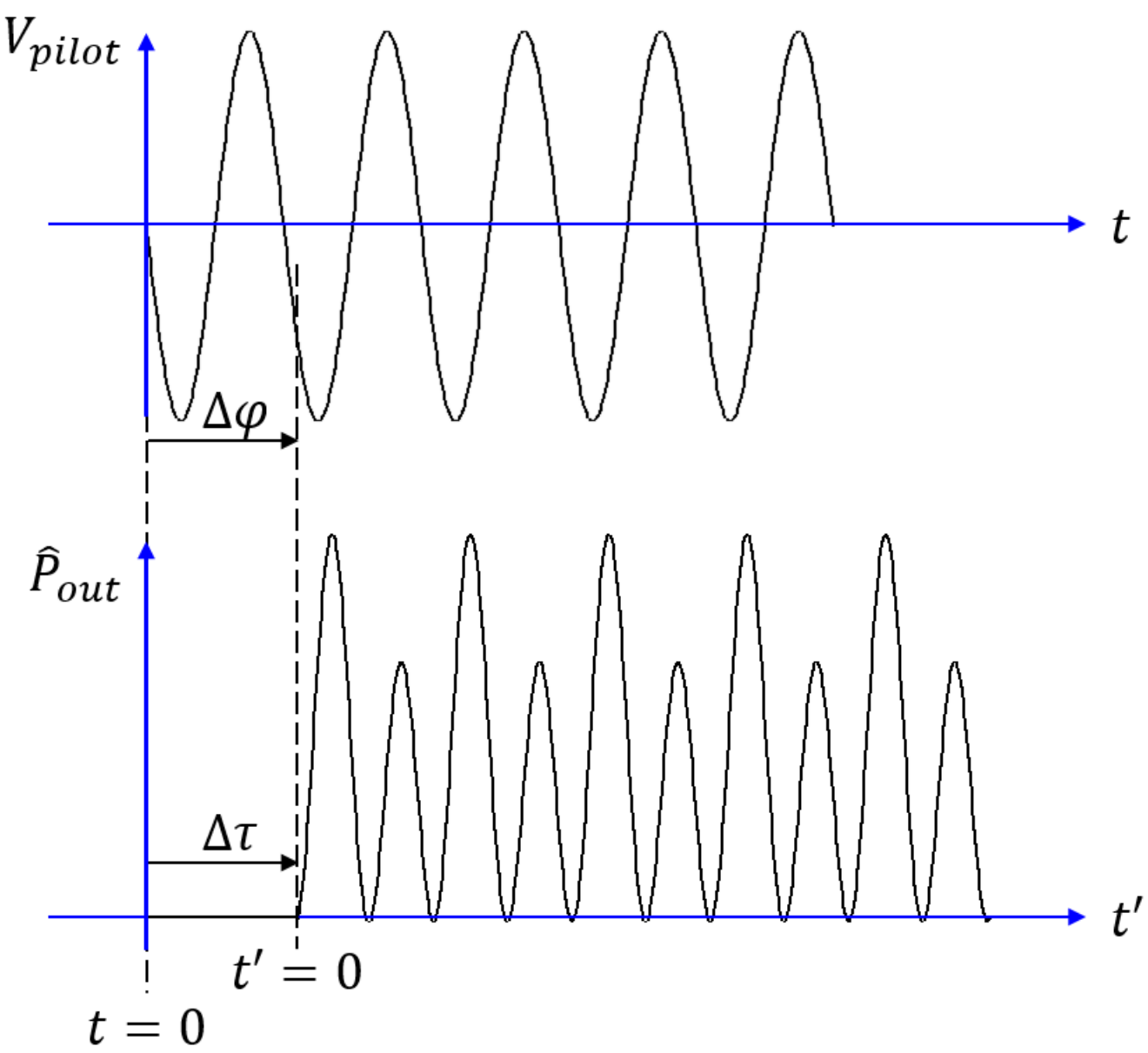

*Figure 2: Temporal relationship between pilot signal $V_{pilot}(t)$ and output power $\hat{P}_{out}(t)$*

In real causal systems, there is always an unavoidable time delay $\Delta\tau > 0$ of $\hat{P}_{out}(t)$ relative to $V_{pilot}(t)$ due to various propagation delays, e.g. by measuring devices, amplifiers, etc., see Figure 2. The output signal to be assigned to the start of the pilot signal at $t = 0$ thus becomes effective in $\hat{P}_{out}$ at $t = \Delta\tau$ resp. $t' = 0$. Since the estimation of $\Delta V$ is based on the evaluation of $\hat{P}_{out}$, its time axis is relevant. With reference to the time axis $t'$ of $\hat{P}_{out}$, the pilot signal $V_{pilot}(t)$ experiences a phase shift of $\Delta\varphi = \Delta\tau \cdot \omega_d$ due to the delay $\Delta\tau$. The effective phase $\varphi_d$ with regard to the time axis of $\hat{P}_{out}$ is thus calculated as follows assuming $\varphi_0 = 0$:

$$\varphi_d = \Delta\tau \cdot \omega_d \,. \tag{13}$$

There are two options regarding the computation of $\Delta V$: $\varphi_d$ can be computed based on $\Delta\tau$, which can e.g. be obtained by measurement of the propagation delay in a real system. Alternatively, the delay $\Delta\tau$ can be estimated and compensated by applying the time axis $t'$ for $\hat{P}_{out}$, i. e. $t' = 0$ corresponds to $t = \Delta\tau$. In this case we obtain $\varphi_d = 0$. In the following, we use the latter approach. With $\varphi_d = 0$ Eq. (11) modifies to:

$$\frac{\boldsymbol{\mathcal{F}}^{+}_{1f2f}(\omega_d)}{\boldsymbol{\mathcal{F}}^{+}_{1f2f}(2\omega_d)} = \frac{S_{1f_d,T_2} \cdot e^{i\frac{\pi}{2}}}{S_{2f_d,T_2}}. \tag{14}$$

As $S_{1f_d,T_2}, S_{2f_d,T_2}$ are real valued it shows that the numerator of Eq. (14) is purely imaginary while the denominator is purely real valued. Accordingly, we can write

$$\frac{imag\left(\boldsymbol{\mathcal{F}}^{+}_{1f2f}(\omega_d)\right)}{real\left(\boldsymbol{\mathcal{F}}^{+}_{1f2f}(2\omega_d)\right)} = \frac{S_{1f_d,T_2}}{S_{2f_d,T_2}} = 4 \cdot \frac{\Delta V}{V_d}. \tag{15}$$

Finally, we obtain

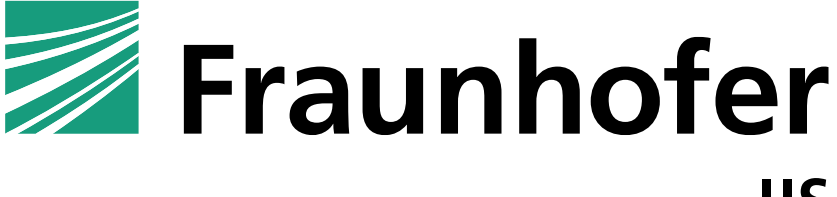

$$\Delta V = \frac{imag\left(\boldsymbol{\mathcal{F}}^{+}_{1f2f}(\omega_d)\right)}{real\left(\boldsymbol{\mathcal{F}}^{+}_{1f2f}(2\omega_d)\right)} \cdot \frac{V_d}{4}. \qquad (16)$$

## 4.2 Discrete-time representation of $P_{out}$

Eq. (16) is based on the evaluation of the time-continuous Fourier transform of the optical EOM output power $P_{out}(t)$. Based on the following assumptions we can replace the time-continuous Fourier transform $\boldsymbol{\mathcal{F}}\{P_{out}(t)\}$ by the Discrete Fourier Transform $\boldsymbol{DFT}\{P_{out}[k]\}$:

- The signal $P_{out}(t)$ is bandwidth limited at $f_{max}$ and sampled equidistantly in compliance with the sampling theorem, i.e. the sampling frequency $f_s = 1/T_s$ is larger than $2 \cdot f_{max}$. As we only consider the two tones at frequencies $f_d$ and $2 \cdot f_d$, we obtain $f_{max} = 2 \cdot f_d$.

- The sequence $P_{out}[k]$ has $N_{DFT}$ elements which comprise an integer multiple of full sine wave periods at frequency $f_d$.

Defining the Discrete Fourier Transform of a sequence $x$ as

$$\boldsymbol{DFT}\{x\} = \frac{1}{N_{DFT}} \cdot \sum_{n=0}^{N_{DFT}-1} x[n] \cdot e^{-i\frac{2\pi}{N_{DFT}}kn}, \qquad (17)$$

we can rewrite Eq. (16) for the time-discrete representation of $P_{out}$ according to

$$\Delta V = \frac{imag\left(\boldsymbol{DFT}\{P_{out}\}\left[z_{1\omega_d}\right]\right)}{real\left(\boldsymbol{DFT}\{P_{out}\}\left[z_{2\omega_d}\right]\right)} \cdot \frac{V_d}{4}. \qquad (18)$$

$z_{1\omega_d}$ indicates the element of the Discrete Fourier Transform which corresponds to the angular frequency $\omega_d$ in the continuous frequency range and, correspondingly, the index $z_{2\omega_d}$ corresponds to the angular frequency $2\omega_d$ .

For further analysis we normalize $\Delta V$ to $V_\pi$:

$$\Delta V_{norm} := \frac{\Delta V}{V_\pi} = \frac{imag\left(\boldsymbol{DFT}\{P_{out}\}\left[z_{1\omega_d}\right]\right)}{real\left(\boldsymbol{DFT}\{P_{out}\}\left[z_{2\omega_d}\right]\right)} \cdot \frac{F}{4} \qquad (19)$$

with

$$F = \frac{V_d}{V_\pi}. \qquad (20)$$

Thus $F$ represents the pilot signal amplitude normalized to $V_\pi$.

## 4.3 Measurement of $\boldsymbol{P_{out}}$ by a photodetector

In a practical system the EOM optical output power is typically measured by a photodetector, providing a voltage, which would in ideal case be proportional to the optical power at EOM output, i.e.

$$V_{out,ideal}(t) = C \cdot P_{out}(t), \tag{21}$$

where $C$ is the overall opto-electric conversion factor (including optical splitter, amplifier etc.) of the photodetector plus amplifier in $V/W$ (volts per watt). Photodetector and amplifier noise are taken into account by additive white Gaussian noise (AWGN) $n(t)$, resulting in $\hat{V}_{out}(t)$ according to

$$\hat{V}_{out}(t) = V_{out,ideal}(t) + n(t) = C \cdot P_{out}(t) + n(t), \tag{22}$$

where $n(t)$ is assumed to have constant power spectral density $S_0$, expressed in $V^2/Hz$. We further assume that the time-continuous signal $\hat{V}_{out}(t)$ is filtered by an ideal rectangular low-pass filter with cut-off frequency $f_s/2$ and then sampled at sampling frequency $f_s$. This results in the sampled sequence

$$\hat{V}_{out}[k] = C \cdot P_{out}[k] + n[k], \tag{23}$$

where the elements $n[k]$ denote uncorrelated samples of a zero-mean Gaussian random sequence with variance $\sigma_n^2 = S_0 \cdot \frac{f_s}{2}$.

Considering the above model of EOM and photodetector operation we can rewrite $\Delta V_{norm}$ given by Eq. (19) according to

$$\Delta V_{norm} = \frac{C \cdot imag\left(\boldsymbol{DFT}\{P_{out}\}[z_{1\omega_d}]\right)}{C \cdot real\left(\boldsymbol{DFT}\{P_{out}\}[z_{2\omega_d}]\right)} \cdot \frac{F}{4} = \frac{imag\left(\boldsymbol{DFT}\{V_{out,ideal}\}[z_{1\omega_d}]\right)}{real\left(\boldsymbol{DFT}\{V_{out,ideal}\}[z_{2\omega_d}]\right)} \cdot \frac{F}{4} \tag{24}$$

using the idealized noise-free photodetector output $V_{out,ideal}$.

When estimating $\Delta V_{norm}$ we only have access to the noisy photodetector samples $\hat{V}_{out}[k]$. For the estimate $\Delta V_{norm,est}$ we obtain

$$\Delta V_{norm,est} = \frac{imag\left(\boldsymbol{DFT}\{\hat{V}_{out}\}[z_{1\omega_d}]\right)}{real\left(\boldsymbol{DFT}\{\hat{V}_{out}\}[z_{2\omega_d}]\right)} \cdot \frac{F}{4}. \tag{25}$$

Exploiting the linearity of the $\boldsymbol{DFT}$ operation we can split up both numerator and denominator into a deterministic summand $\boldsymbol{DFT}\{V_{out,ideal}\}[z_{*\omega_d}]$ and a noise component $\boldsymbol{DFT}\{n\}[z_{*\omega_d}]$:

$$\Delta V_{norm,est} = \frac{F \cdot imag\left(\boldsymbol{DFT}\{V_{out,ideal}\}[z_{1\omega_d}] + \boldsymbol{DFT}\{n\}[z_{1\omega_d}]\right)}{4 \cdot real\left(\boldsymbol{DFT}\{V_{out,ideal}\}[z_{2\omega_d}] + \boldsymbol{DFT}\{n\}[z_{2\omega_d}]\right)}. \tag{26}$$

## 4.4 Estimation error

In the following we compute the estimation error of $\Delta V_{norm,est}$, i.e. its deviation from $\Delta V_{norm}$ due to photodetector noise, quantified by $S_0$.

The computation of $\Delta V_{norm,est}$ according to Eq. (26) is a quotient of the numerator $F \cdot imag(\ldots)$ and denominator $4 \cdot real(\ldots)$. The Discrete Fourier Transform is calculated in both numerator and denominator from the superposition of a deterministic signal $V_{out,ideal}$ plus zero-mean white Gaussian noise $n$. The DFT of $V_{out,ideal}$ only depends on the system parameters and the value of $\Delta V$. It has no stochastic component and therefore results in a fixed value. The DFT of zero mean white Gaussian noise results in a zero mean white Gaussian random sequence.

Therefore, numerator ("$num$") and denominator ("$den$") of Eq. (26) can be considered as two uncorrelated discrete normally distributed random variables defined by $N_{num}(\mu_{num}, \sigma_{num}^2)$ and $N_{den}(\mu_{den}, \sigma_{den}^2)$, where $N(\mu, \sigma^2)$ defines a normal distribution. $\mu_{num}, \mu_{den}$ are the respective mean values for the numerator and denominator and $\sigma_{num}^2, \sigma_{den}^2$ the corresponding variances. The mean values $\mu_{num}, \mu_{den}$ are determined exclusively by the DFT of $V_{out,ideal}$, whereas the variances $\sigma_{num}^2, \sigma_{den}^2$ are exclusively determined by the DFT of the noise sequence $n$.

First, the four parameters of the normal distributions in the numerator and denominator are calculated, then the variance of the estimation error of $\Delta V_{norm,est}$ is determined.

### Variances $\sigma_{num}^2, \sigma_{den}^2$

$\sigma_{num}^2$ and $\sigma_{den}^2$ are obtained by considering the DFT of the noise sequence $n[k], k \in \{0, 1, \ldots, N_{DFT} - 1\}$, where $N_{DFT}$ is the DFT length. The DFT based on the definition according to Eq. (17) of a zero-mean white Gaussian noise process $N(0, \sigma_n^2)$ is in turn normally distributed with the variance $\sigma_{n,DFT}^2 = \frac{\sigma_n^2}{N_{DFT}}$, i. e. the variance is reduced by the factor $N_{DFT}$. Considering the factor $F/4$ and the $real(\ldots)$ and $imag(\ldots)$ operations halving the effective noise we obtain for the variances of numerator and denominator:

$$\sigma_{num}^2 = \frac{F^2 \cdot \sigma_n^2}{2 \cdot N_{DFT}} = \frac{F^2 \cdot S_0 \cdot f_s}{4 \cdot N_{DFT}} \tag{27}$$

$$\sigma_{den}^2 = \frac{4^2 \cdot \sigma_n^2}{2 \cdot N_{DFT}} = \frac{4 \cdot S_0 \cdot f_s}{N_{DFT}}. \tag{28}$$

With $T_d = N_{DFT}/f_s$ representing the time duration of the pilot signal processed by the DFT we obtain

$$\sigma_{num}^2 = \frac{F^2 \cdot S_0}{4 \cdot T_d} \tag{29}$$

$$\sigma_{den}^2 = \frac{4 \cdot S_0}{T_d}. \tag{30}$$

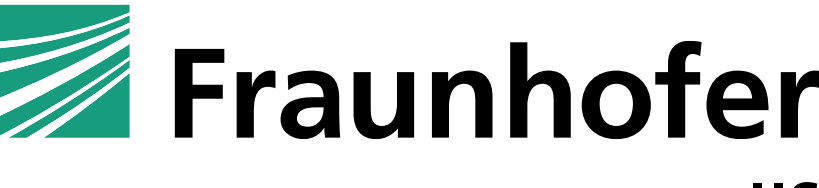

## Mean values $\mu_{num}, \mu_{den}$

Regarding the mean values we have to consider the DFT of $V_{out,ideal} = C \cdot P_{out}$ at the indexes $z_{1\omega_d}$ (numerator) and $z_{2\omega_d}$ (denominator).

According to the DFT definition of Eq. (17) the DFT of a pure sine function $x[n] = sin(2\pi n \cdot f_d/f_s)$ for $\omega_d = 2\pi f_d$ and $n \in \{0, 1, \dots, N_{DFT} - 1\}$ results in $\boldsymbol{DFT}\{x\}[z]$ with

$$\boldsymbol{DFT}\{x\}[z_{\omega_d}] = -0.5j \quad \text{for } z_{\omega_d} = N_{DFT}\frac{f_d}{f_s}, \tag{31}$$

assuming that $N_{DFT}$ comprises an integer multiple of sine wave periods.

Recalling the computation of $P_{out}$ according to Eqs. (8), (9) with $\varphi_d = 0$ we obtain:

$$\boldsymbol{DFT}\{\hat{V}_{out,ideal}\}[z_{1\omega_d}] = -\frac{1}{2}j \cdot C \cdot P_{in} \cdot \frac{f_{ib}}{2}\left(\frac{\pi}{V_\pi}\right)^2 \cdot 2 \cdot \Delta V \cdot V_d, \tag{32}$$

$$\boldsymbol{DFT}\{\hat{V}_{out,ideal}\}[z_{2\omega_d}] = -\frac{1}{2} \cdot C \cdot P_{in} \cdot \frac{f_{ib}}{2}\left(\frac{\pi}{V_\pi}\right)^2 \cdot \frac{1}{2}(V_d)^2. \tag{33}$$

For the mean values of numerator and denominator we finally obtain:

$$\mu_{num} = -\frac{\pi^2}{2} \cdot C \cdot P_{in} \cdot f_{ib} \cdot F^2 \cdot \frac{\Delta V}{V_\pi}, \tag{34}$$

$$\mu_{den} = -\frac{\pi^2}{2} \cdot C \cdot P_{in} \cdot f_{ib} \cdot F^2. \tag{35}$$

## Estimation error of $\Delta V_{norm,est}$

The probability distribution function of $\Delta V_{norm,est}$ is a ratio distribution of two independent normally distributed variables, as described above. As both numerator and denominator of Eq. (26) have non-zero mean values it is denoted as uncorrelated non-central normal ratio. Although the associated probability density function (PDF) can be precisely calculated, it results in a comparatively complicated formula, see e. g. [11]. It shows that the quotient of two independent normal distributions does not necessarily result in a normal distribution again. However, under certain conditions a normal approximation of the ratio distribution's PDF is possible, see [12]. In simple terms, the smaller the variance of the denominator $\sigma_{den}^2$ in relation to the squared mean value of the denominator $\mu_{den}^2$, the better the approximation.

According to the referred approximation [12] the estimation error variance of $\Delta V_{norm,est}$ is given by

$$\sigma_{\Delta V,norm,est}^2 \approx \frac{\mu_{num}^2}{\mu_{den}^2} \cdot \left(\frac{\sigma_{num}^2}{\mu_{num}^2} + \frac{\sigma_{den}^2}{\mu_{den}^2}\right). \tag{36}$$

By inserting $\sigma_{num}^2, \sigma_{den}^2, \mu_{num}, \mu_{den}$ from Eqs. (29), (30), (34), (35) into Eq. (36) we finally obtain the following result:

$$\sigma^2_{\Delta V,norm,est} \approx \frac{1}{T_d} \cdot \frac{S_0}{C^2} \cdot \frac{F^2 + 16 \cdot \Delta V^2_{norm}}{(\pi^2 \cdot P_{in} \cdot f_{ib} \cdot F^2)^2}. \quad (37)$$

It shows that the estimation error variance $\sigma^2_{\Delta V,norm,est}$ is a product of several terms:

- $\frac{1}{T_d}$: The variance inversely scales with $T_d$, which is the length of the pilot signal used as input to the DFT.

- $\frac{S_0}{C^2}$: This ratio represents the parameters of the photodetector, where $S_0$ is the noise power spectral density and $C$ the opto-electric conversion factor. As expected, a small noise power and a large conversion factor should be aimed for.

- $\frac{F^2+16\cdot\Delta V^2_{norm}}{(\pi^2\cdot P_{in}\cdot f_{ib}\cdot F^2)^2}$: This ratio basically depends on the optical input power $P_{in}$, the relative amplitude of the pilot signal $F = V_d/V_\pi$, and the variable to be estimated $\Delta V_{norm}$. The latter is significant insofar as the estimation error increases with the squared amount of the variable to be estimated.

It should also be noted that the estimation error variance is independent of the pilot signal frequency $f_d$, which provides a degree of freedom in the design of a real-word system.

## 4.5 Summary of the assumptions made

Before results are presented in the following section, all the assumptions and approximations made so far are summarized here in compact form.

**EOM model:**

A. EOM opto-electrical transfer function modeled by Eq. (1), no parameter drift within $T_d$ (=time period covered by DFT input sequence).

B. 2nd order polynomial approximation of the sinusoidal EOM transfer function in the vicinity of considered minimum (Eq. (7)).

**Photodetector model:**

C. Completely linear photodetector, adding white Gaussian noise (AWGN) to output voltage.

D. Photodetector output ideally low pass filtered and sampled in compliance with sampling theorem.

E. Propagation delay $\Delta\tau$ in photodetector path known and fully compensated, i.e. $\varphi_d = 0$.

**Estimation error variance:**

F. Variance of the uncorrelated non-central normal ratio approximated by Eq. (36).

# 5 Results

## 5.1 Estimation error

In this section the estimation error for $\Delta V_{norm,est}$ calculated according to Eq. (37) is compared to corresponding simulation results. The latter are based on simulated EOM output power, followed by the ideal photodetector plus AWGN. This means that in the simulation the assumptions B) and F) of section 4.5 are not applied, i.e. Eq. (16) is applied to the original EOM model (without 2[nd] order polynomial approximation).

**Model parameters**

A laser beam with constant power of $1\,W$ is assumed as input to the EOM. For the EOM the optical imbalance factor was set to $f_{ib} = 0.5$.

With regard to the photodetector, parameters were chosen that are in line with commercially available photodiodes. A photodiode conversion factor of 10 $V/W$ combined with an optical splitter that couples 1 % of the EOM output to the photodetector results in an assumed opto-electrical conversion factor of $C = 0.1\ V/W$. The noise power spectral density was modeled as $S_0 = \left(50\ pV/\sqrt{Hz}\right)^2$. The photodetector output was sampled at $f_s = 5\ MHz$.

The applied pilot signal has a frequency of $f_d = 0.5\,MHz$ and an active duration (used as DFT input) of $T_d = 50\,\mu s$, corresponding to 25 sine wave periods.

**Results and discussion**

Figure 3 shows the standard deviation of the estimation error $\sigma_{\Delta V,norm,est}$ over the normalized pilot signal amplitude $F = V_d/V_\pi$.

Black lines indicate simulation results and red lines the corresponding calculated results. Regarding the normalized offset of the current operating point $\hat{V}_{min}$ to the actual minimum of the EOM transfer function $V_{min}$, i. e.$\Delta V/V_\pi = \left(V_{min} - \hat{V}_{min}\right)/V_\pi$, two cases are considered: $\Delta V/V_\pi$ =0 (solid lines with asterisks) and $\Delta V/V_\pi = 0.002$ (dashed lines with circles).

First focusing on the calculated results (red) we see that the standard deviation of the estimation error decreases with increasing pilot signal amplitude. This is expected as the photodetector noise is constant whereas the EOM output increases with increasing pilot signal power. It can also be observed that an offset of $\Delta V/V_\pi = 0$, i.e. the pilot signal is applied on top of the actual minimum voltage $V_{min}$, results in a better estimation accuracy than for $\Delta V/V_\pi = 0.002$. Even in the latter case a standard deviation $\sigma_{\Delta V,norm,est} < 10^{-4}$ can be achieved for normalized pilot signal amplitudes $F$ larger than approx. $10^{-3}$.

For better illustration we consider the following example: From a previous estimate of $\hat{V}_{min}$ there is still an offset (mismatch) to the actual $V_{min}$ of $\left(V_{min} - \hat{V}_{min}\right)/V_\pi = \Delta V/V_\pi = 0.002$. At this operating point and assuming a pilot signal amplitude of e.g. $F = 10^{-3}$ we obtain $\sigma_{\Delta V,norm,est} \approx 10^{-4}$. For a normal distribution this means that the next estimate will be at a probability of 99.7 % within the $\pm 3$-sigma range, which is approximately $3 \cdot 10^{-4}$. So the magnitude of the next offset $\Delta V/V_\pi$ will be very likely reduced from 0.002 to less than 0.0003.

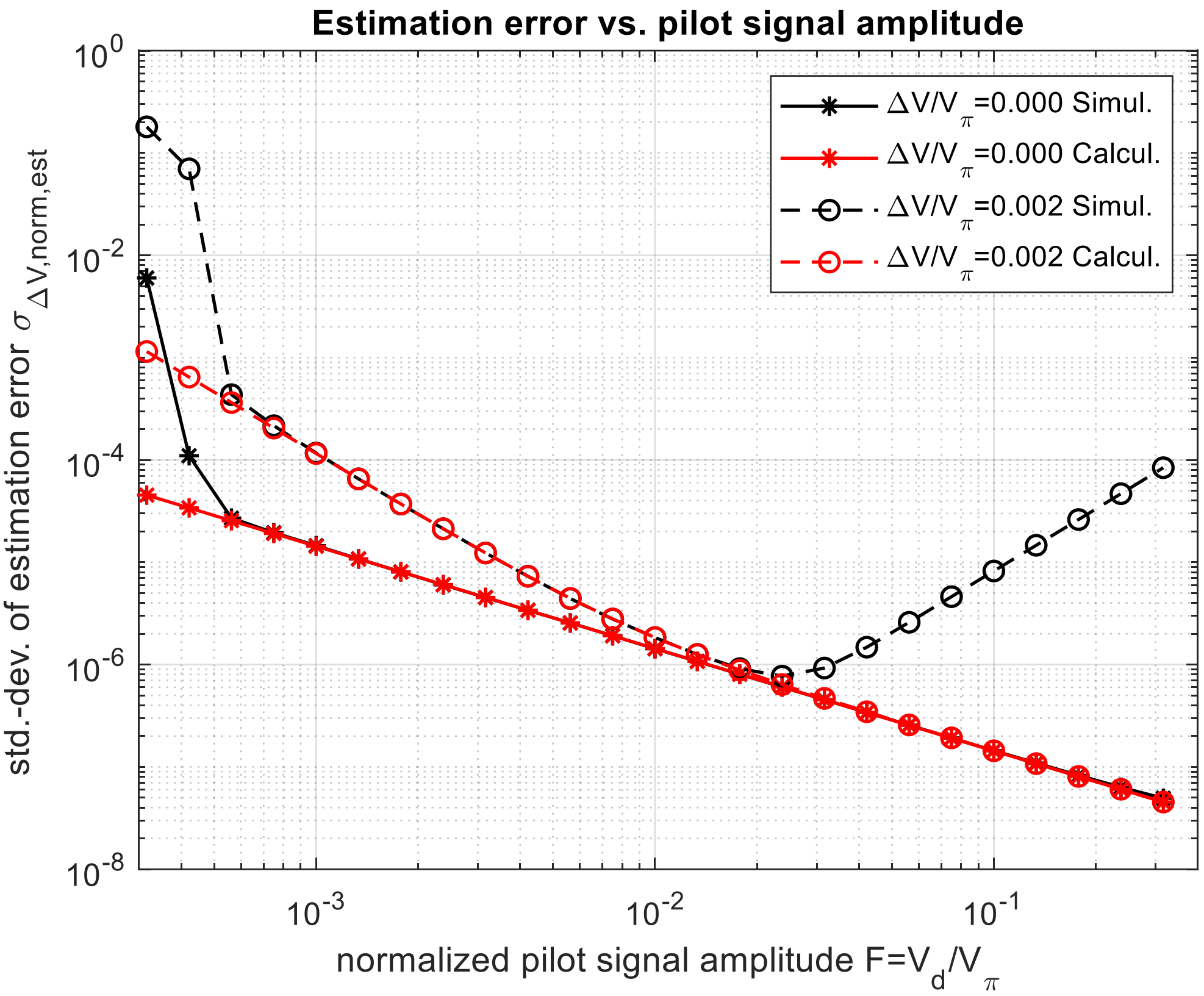

*Figure 3: Standard deviation of estimation error $\sigma_{\Delta V,norm,est}$, simulation and calculation*

A comparison of simulation and calculation shows that the results for very low and high values of $F$ differ from each other. For comparatively small pilot signal amplitudes $F \leq 5 \cdot 10^{-4}$ the simulation results show a significantly larger standard deviation than the calculation. The reason for this is that for a very small pilot signal, the signal-to-noise ratio (SNR) at the photodetector output becomes very small, i.e. the uncorrelated non-central normal ratio of by Eq. (26) can no longer be approximated by a normal distribution. Therefore, assumption F) from section 4.5 is not applicable and the variance approximation described by Eq. (36) is also not valid.

For relatively large values of $F \geq 0.02$ and $\Delta V/V_\pi = 0.002$, an increasing discrepancy between simulation and calculation can also be observed. The reason is that the pilot signal is comparatively large in this region and the 2nd order polynomial approximation of the sinusoidal EOM transfer function in the vicinity of its minima does no longer hold. Assumption B) of section 4.5 is invalid. Moreover, it turns out that further increasing the amplitude of the pilot signal even degrades the estimation performance of $\Delta V_{norm,est}$.

In summary, it can be stated that in a middle range of $F$, in our example between $5 \cdot 10^{-4}$ and 0.02, the theoretical calculation of the estimation error variance agrees very well with the simulation results. At the edges, however, there are deviations between calculation and simulation. The amplitude of the pilot signal should be selected neither too low (poor SNR) nor too high (increase in the estimation error).

## 5.2 Black level with activated pilot signal

In a quantum computer system, it may be very important that the EOM attenuates a constant power laser beam applied to the input as much as possible, i.e. that the minimum output level, also denoted as black level, is as low as possible. This aspect is addressed below in connection with the pilot signal.

In the EOM model according to Eq. (1) the maximum possible EOM output power is equivalent to $P_{in}$ for $f_{ib} = 0.5$. The minimum EOM output without pilot signal is determined by the current operating point $V_{DC} = \hat{V}_{min}$, expressed by the offset $\Delta V$ relative to the actual $V_{min}$. When applying a pilot signal on top of $V_{DC}$ the EOM output power oscillates as shown exemplary in the lower part of Figure 2.

We define the relative black level $P_{bl,rel}$ during the active pilot signal as the maximum of the oscillating EOM output power $P_{out}(t)$ relative to $P_{in}$. In the log domain this results in:

$$P_{bl,rel} := 10 \cdot log_{10}\left(\frac{max\left(P_{out}(t)\right)}{P_{in}}\right). \tag{38}$$

If the sign is inverted this is equivalent to the minimum attenuation of the EOM input power in the presence of the pilot signal.

Figure 4 depicts the relative black level over the amplitude of the pilot signal for $\Delta V/V_{\pi} = 0$ and 0.002.

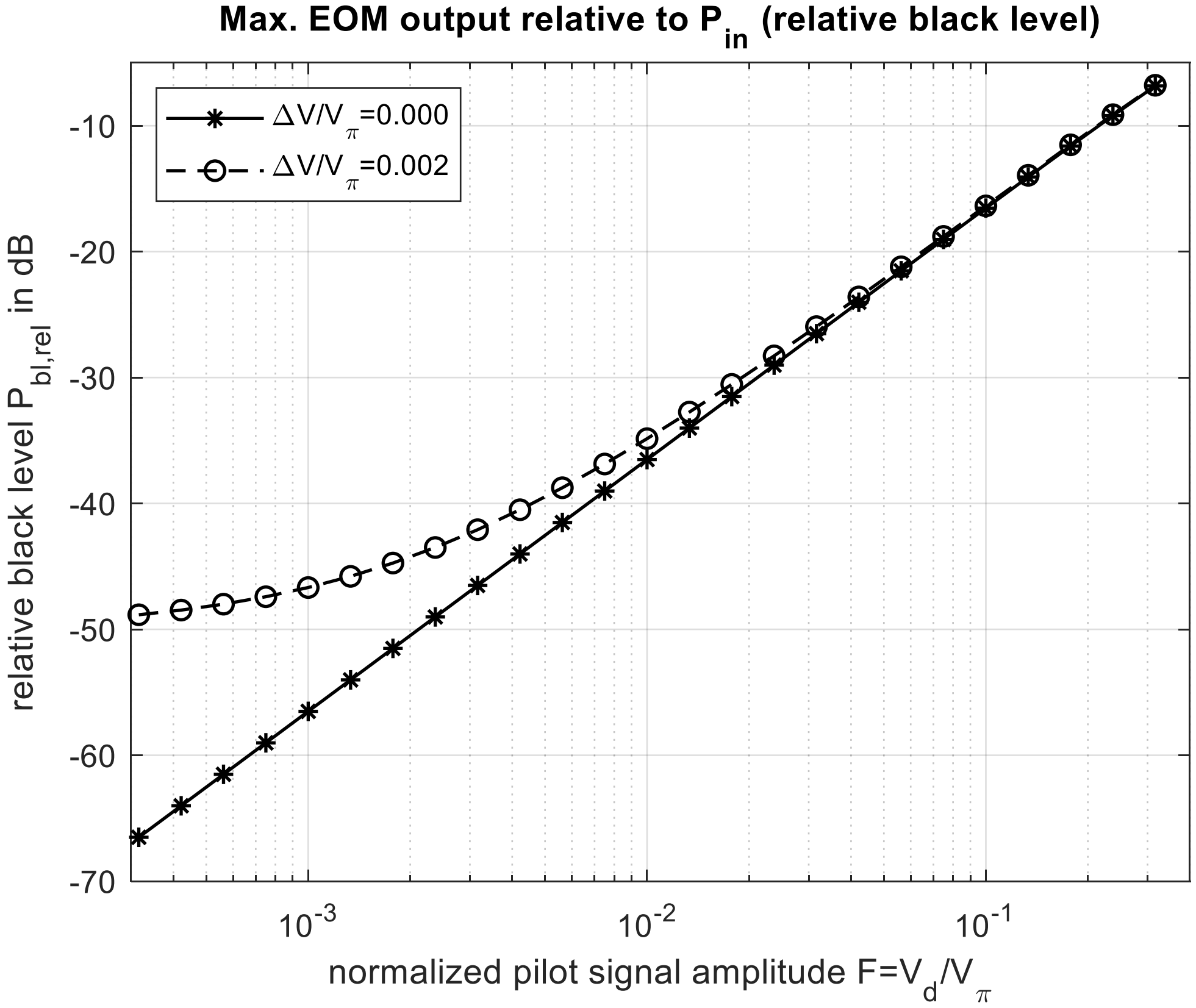

*Figure 4: Relative black level $P_{bl,rel}$ vs. normalized pilot signal amplitude*

For $\Delta V/V_\pi = 0$ the relative black level increases linearly with the pilot signal amplitude in the depicted range. For $\Delta V/V_\pi = 0.002$ it converges towards a lower limit with decreasing $F$, as even without a pilot signal ($F = 0$) the EOM output power has a non-zero value.

## 5.3 Pilot signal design criteria

The algorithm described in this paper allows to estimate the offset $\Delta V$ of the current DC operating point $V_{DC} = \hat{V}_{min}$ from the actual minimum output voltage $V_{min}$ based on a pilot signal. This estimated offset can be used as input to a black level control algorithm that intends to keep the EOM optical output as small as possible. Various kinds of control algorithms may be applied, the description of which is not the subject of this paper.

The pilot signal is defined by the following parameters: amplitude $V_d$, duration $T_d$, frequency $f_d$, and repetition rate. The design criteria for the pilot signal are discussed below in a qualitative way.

**Amplitude $V_d$**

The amplitude has impact both on the estimation accuracy and on the EOM black level, which is increased by the pilot signal itself. During periods when the EOM output is blocked from the subsequent quantum computing system by an additional device (e. g. shutter) or when the quantum computer is not sensitive to light, the pilot signal amplitude may be increased. This allows for a quick and accurate estimation of the offset $\Delta V$. However, as shown in Figure 3, $V_d$ should not exceed a certain level beyond which the estimation accuracy starts to decrease. In periods when the smallest possible black level is crucial, the pilot signal amplitude shall be kept small. Estimation accuracy may be maintained by a longer duration $T_d$ of the pilot signal.

**Duration $T_d$**

As shown above the duration of the pilot signal fed into the DFT proportionally increases the estimation quality. In a practical system this duration may be limited by the following aspects: The black level voltage $V_{min}$ of the EOM should not drift within the duration of the pilot signal. In addition, the quantum computing system may impose a restriction on the pilot signal duration, e. g. due to time intervals between subsequent pulses in a pulse train.

**Frequency $f_d$**

Beside restrictions mentioned above regarding integer multiples of pilot signal periods in the DFT input sequence, the pilot signal frequency $f_d$ theoretically has no impact on the estimation accuracy. However, there are some practical aspects to be considered. We have assumed that the phase angle $\varphi_d$ can be set to 0 by compensating the propagation delay $\Delta\tau$, see section 4.1. Estimation errors of $\Delta\tau$ result in a corresponding phase error which linearly increases with $f_d$, see Eq. (13). Furthermore, $f_d$ should be chosen in a reasonable order of magnitude w.r.t the frequency responses of the EOM, the photodetector and the subsequent ADC. These aspects suggest that the pilot signal frequency should be reasonably low.

**Repetition rate**

The proposed evaluation of the EOM output associated with the pilot signal is block-based. It therefore makes sense to only apply the pilot signal when an estimation of $\Delta V$ is actually to be performed, so that the pilot signal may have a burst-like characteristic. The time interval between two pilot signal bursts depends firstly on the speed at which the minimum output voltage $V_{min}$ of the EOM drifts and secondly on the acceptable estimation error. The type of control loop (e.g. predictive or non-predictive) that processes the estimated values $\Delta V_{est}$ and accordingly updates $\hat{V}_{min}$ is also relevant.

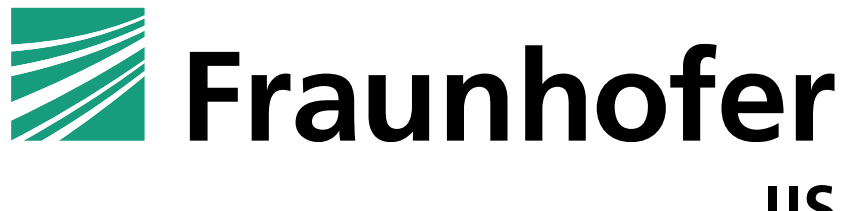

# 6 Conclusion

We have proposed a method to estimate the EOM minimum point control voltage $V_{min}$ based on a small sinusoidal pilot signal, which is applied to the EOM control input. Specifically, the algorithm estimates the voltage offset $\Delta V$ of the current operation point $\hat{V}_{min}$ relative to the actual minimum output voltage $V_{min}$ of the EOM. These estimates can be used as input to a control loop aimed at minimizing the EOM's optical output power. The presented approach is considered in particular for use in a quantum computing system in which a continuous wave (CW) laser is modulated by an EOM to generate both optical pulses or pulse trains and periods with shuttered output.

The computation of the estimated values for $\Delta V$ is derived theoretically, initially for an infinite length of the pilot signal. If the timing between the pilot signal and the corresponding feedback signal from the photodetector coupled to the EOM output is known, both magnitude and sign of $\Delta V$ can be estimated jointly. The proposed algorithm has the advantage that the estimation of $\Delta V$ is independent of physical EOM parameters such as e.g. the half-wave voltage $V_\pi$ or the optical imbalance factor.

Subsequently, we consider some relevant aspects of a real-world implementation, such as e.g. converting and sampling the optical power of the EOM output with a noisy photodetector and processing sampled sequences of finite duration. On this basis, and with a simplifying assumption, the variance of the estimation error is calculated theoretically. This makes it possible to quantitatively evaluate the effects of photodetector parameters (spectral noise power density, electro-optical conversion factor) and parameters of the pilot signal (amplitude, duration) on the estimation accuracy.

Simulations show that the calculated estimation error variance agrees very well with corresponding simulation results in a wide range. For comparatively very small and large pilot signals, deviations between simulation and calculation are observed, which can be attributed to the violation of modelling assumptions. It turns out that from a certain point onwards, a further increase of the pilot signal even degrades the estimation accuracy. Additional simulation results show the impact of the pilot signal on the achievable minimum EOM output power, i.e. the increase of the black level due to the applied pilot signal itself.

Finally, design criteria for the required pilot signal in terms of its amplitude, duration, frequency and repetition rate are discussed in particular w.r.t. the application of the EOM in a quantum computing system.

## Acknowledgements

The research is part of the Munich Quantum Valley, which is supported by the Bavarian state government with funds from the Hightech Agenda Bayern Plus.